\documentclass[10pt]{iopart}
\usepackage{graphicx}
\usepackage{xcolor}
\usepackage{dcolumn}% Align table columns on decimal point
\usepackage{bm}% bold math
\usepackage{braket}
\usepackage[normalem]{ulem}
\usepackage{siunitx}
\usepackage{upgreek}

\begin{document}
%%% article in English
%\lat

%%% declaration of a new mathematical operator
%\DeclareMathOperator{\sign}{sign}

%%% article title
\title{Compact laser system with frequency stability dissemination for optical clocks and quantum computing.}

%%% article title - for colontitle (at the top of the page)
%\rtitle{Compact laser system for frequency stability distribution \ldots}

%%% article title - for the table of contents (usually identical with \title)
%\sodtitle{Compact laser system for frequency stability distribution in problems of metrology and quantum computing}

\author{
M.I. Shakirov$^{1}$, K.S. Kudeyarov$^{1}$, N.O. Zhadnov$^{1}$, D.S. Kryuchkov$^{1}$, A.V. Tausenev$^{1}$, K.Yu. Khabarova$^{1}$, N.N.~Kolachevsky $^{1,2}$}
%%% author(s) - for colontitle (at the top of the page)
%\rauthor{M.I. Shakirov, K. S. Kudeyarov, N. O. Zhadnov, D. S. Kryuchkov, A. V. Tausenev, K. Yu. Khabarova, N. N. Kolachevsky }

%%% author(s) - for table of contents
%\sodauthor{M.I. Shakirov, K. S. Kudeyarov, N. O. Zhadnov, D. S. Kryuchkov, A. V. Tausenev, K. Yu. Khabarova, N. N. Kolachevsky }

%%% author's address(es)
\address{$^1$Lebedev Physical Institute, Russian Academy of Sciences, 119991, Moscow, Russia ; \\
$^2$Russian Quantum Center, Skolkovo, 121205, Moscow, Russia}
\ead{m.shakirov@lebedev.ru}
%%% dates of submition & resubmition (if submitted once, second argument is *)
\begin{indented}
\item[] April 2025
\end{indented}

\begin{abstract}
Modern experiments in quantum metrology, sensing, and quantum computing require precise control of the state of atoms and molecules, achieved through the use of highly stable lasers and microwave generators with low phase noise. One of the most effective methods for ensuring high frequency stability is stabilization using a high-finesse Fabry-Pérot reference cavity. However, implementing separate stabilization systems for each laser increases the complexity and size of the setup, limiting its use to laboratory conditions. An alternative approach is the use of a femtosecond optical frequency comb, which transfers the noise characteristics of a single stabilized frequency reference to other wavelengths in the optical and microwave ranges.
In this work, we demonstrate a scheme for transferring frequency stability from an ultrastable laser at 871 nm to a laser at 1550 nm.
Measurements using the three-cornered hat method show that the stabilized laser exhibits a fractional frequency instability of less than $4\times10^{-15}$ for averaging times between 0.4 and 2 s, and below $10^{-14}$ for intervals ranging from 0.2 to 500 s. The femtosecond optical frequency comb and the cavity-stabilized laser were designed to meet compactness and portability requirements to enable field and onboard applications.
%Современные эксперименты в области квантовой метрологии, сенсорики и квантовых вычислений требуют прецизионного контроля состояния атомов и молекул, что достигается за счет использования высокостабильных лазерных и микроволновых генераторов с низким уровнем фазовых шумов. Одним из наиболее эффективных методов обеспечения высокой стабильности частоты является стабилизация по моде высокодобротного опорного резонатора Фабри-Перо. Однако создание отдельных систем стабилизации для каждого лазера увеличивает сложность и габариты установки и ограничивает её применение лабораторными условиями. Альтернативным подходом является использование фемтосекундной гребенки оптических частот, которая позволяет передавать шумовые характеристики сигнала частоты единственного стабилизированного источника на другие длины волн в оптическом и микроволновом диапазонах.
%В данной работе реализована схема передачи стабильности частоты от ультрастабильного лазера с длиной волны 871 нм к лазеру на длине волны 1550 нм.
%Результаты измерений методом треуголки показали, что стабилизированный лазер демонстрирует относительную нестабильность частоты менее $4\times10^{-15}$ для интервалов от 0,4 до 2 с и менее $10^{-14}$ для интервалов усреднения от 0,2 до 500 с. Фемтосекундная гребенка и лазер с опорным резонатором разработаны с учетом требований компактности и портативности для использования в полевых и бортовых задачах.
\end{abstract}

%%% PACS numbers
%\PACS{74.50.+r, 74.80.Fp}

\maketitle
\ioptwocol
\section{Introduction}
%Появление ультрастабильных лазерных систем со спектральной шириной линии излучения порядка 1 Гц \cite{young1999visible} сделало возможной прецизионную спектроскопию сильно запрещенных атомных переходов. В сочетании с фемтосекундными синтезаторами оптических частот \cite{holzwarth2000optical} это стало ключевым шагом в развитии эры оптических атомных часов.
The development of ultrastable laser systems with spectral linewidths on the order of 1 Hz \cite{young1999visible} enabled precision spectroscopy of highly forbidden atomic transitions. Combined with femtosecond optical frequency synthesizers \cite{holzwarth2000optical}, this marked a key step at the beginning of the era of optical atomic clocks.
%Лучшие современные ультрастабильные лазерные системы демонстрируют выдающиеся характеристики: относительная нестабильность частоты излучения составляет менее $10^{-16}$ на временах усреднения $0.1-10^{3}\ $с \cite{kedar2023frequency,Matei2017a, Hafner2015}, и их параметры продолжают улучшаться. Такой уровень достигается благодаря стабилизации лазера по собственной моде вакуумного высокодобротного (c резкостью $>10^{5}$) резонатора Фабри-Перо методом Паунда-Дривера-Холла (ПДХ) \cite{Drever1983}. Область применения таких лазеров весьма широка -- излучение с низкими фазовыми шумами, высокой стабильностью частоты и большой длиной когерентности востребовано во многих фундаментальных и прикладных задачах, например, детектировании гравитационных волн \cite{capote2025advanced}, поиске дрейфа фундаментальных констант \cite{takamoto2015frequency}, поиске темной материи \cite{savalle2021searching}, построении карты гравитационного потенциала Земли \cite{abich2019orbit}. 
State-of-the-art ultrastable laser systems exhibit exceptional performance: their fractional frequency instability is below $10^{-16}$  for averaging times of  $0.1-10^{3}$ s \cite{kedar2023frequency,Matei2017a,Hafner2015}, and their parameters continue to improve. This level of stability is achieved by stabilizing the laser to a mode of a vacuum high-finesse ($>10^{5}$) Fabry-Pérot cavity using the Pound-Drever-Hall (PDH) technique  \cite{Drever1983}. The applications of such lasers are vast — radiation with low phase noise, high frequency stability, and long coherence length is critical for numerous fundamental and applied tasks, including gravitational wave detection \cite{capote2025advanced}, searches for variations in fundamental constants \cite{takamoto2015frequency}, dark matter detection \cite{savalle2021searching}, and mapping Earth’s gravitational potential \cite{abich2019orbit}.

%Наиболее заметным применением ультрастабильных лазерных систем являются современные оптические атомные стандарты частоты и времени, в которых они играют роль ``маховика''. Частота излучения часового лазера привязывается к частоте спектрально узкого ``часового'' перехода ансамбля лазерно-охлажденных атомов \cite{aeppli2024clock} или одиночного иона \cite{brewer2019al+}. При этом метрологические характеристики часового лазера определяют относительную нестабильность частоты системы на временах измерения порядка 1-10 с, пока происходит набор статистики атомного референса, а также эффективность усреднения флуктуаций частоты на более долгих временах. Помимо лазера для спектроскопии часового перехода, во многих экспериментах с оптическими часами необходимы дополнительные высокостабильные лазерные системы, например, для глубокого лазерного охлаждения на узких переходах \cite{provorchenko2023deep,проворченко2024лазерное} или создания оптической решетки.
The most prominent application of ultrastable laser systems is in modern optical atomic clocks, where they serve as a "flywheel". The clock laser frequency is locked to a spectrally narrow "clock" transition in an ensemble of laser-cooled atoms \cite{aeppli2024clock} or a single ion \cite{brewer2019al+}. The metrological performance of the clock laser determines the system fractional  frequency instability over measurement intervals of 1–10 s (during atomic reference data acquisition) and governs the efficiency of averaging out frequency fluctuations over longer timescales. Beyond the clock transition spectroscopy laser, many optical clock experiments require additional highly stable laser systems, for example, for deep laser cooling on narrow transitions \cite{provorchenko2023deep,provorchenko2024laser} or optical lattice trapping.

%В настоящее время наблюдается активный переход от лабораторных оптических часов к транспортируемым системам \cite{koller2017transportable,grotti2024long,takamoto2020test,huang2020geopotential}, что необходимо для расширения области применения оптических часов, проведения экспериментов в полевых условиях. Была продемонстрирована работа на борту космических аппаратов ультрастабильных лазеров \cite{thompson2011flight,abich2019orbit} и фемтосекундной гребенки \cite{yun2023Laser,shao2024optical}. 
Currently, there is an active shift from laboratory optical clocks to transportable systems \cite{koller2017transportable,grotti2024long,takamoto2020test,huang2020geopotential}, which is essential to expand the applications of optical clocks and to enable field experiments. The operation of ultrastable lasers \cite{thompson2011flight,abich2019orbit} and femtosecond frequency combs \cite{yun2023Laser,shao2024optical} aboard spacecraft has already been demonstrated.
%Особенности эксплуатации вне лабораторных условий накладывают на конструкцию часовых лазеров такие требования, как малые размеры и вес, низкая вибрационная чувствительность и повышенная надежность \cite{kraus2025ultra,hill2021dual}.
Operation outside laboratory conditions imposes specific design requirements on clock lasers, including compact size, low weight, reduced vibration sensitivity, and improved reliability \cite{kraus2025ultra,hill2021dual}.

%Лазерные системы с длиной когерентности излучения в десятки тысяч километров могут использоваться для передачи точных сигналов частоты и времени по волоконным \cite{grotti2024long} и открытым \cite{kudeyarov2020frequency} каналам, что открывает новые перспективы для навигации. 
%Кроме того, с их помощью могут решаться и задачи сенсорики -- когерентной рефлектометрии и дальнометрии, релятивистской геодезии и прецизионной интерферометрии. 
%%Moreover, these systems can address sensing applications, including coherent reflectometry, ranging, relativistic geodesy, and precision interferometry.

%Также перспективным направлением для применения стабилизированных лазеров является генерация  микроволновых сигналов, стабильность которых обеспечивается опорным оптическим резонатором. Как и в случае оптических часов, микроволновый сигнал формируется в результате деления частоты лазера при помощи фемтосекундной гребенки \cite{fortier2011generation,xie2017photonic,киреев2020синтезатор,шелковников2023фотонный}. 
Another promising application of stabilized lasers is the generation of microwave signals whose stability is provided by an optical reference cavity. Similarly to optical clocks, the microwave signal is produced through laser frequency division using a femtosecond frequency comb \cite{fortier2011generation,xie2017photonic,kireev2020radiofrequency,shelkovnikov2023photonic}.

%Кроме выше перечисленного в последнее время появился новый класс задач, где спектральные характеристики лазерного источника оказываются критически важными. К таким задачам, в частности, относятся квантовые вычисления на холодных атомах в оптических пинцетах \cite{levine2018high} или лазерно-охлажденных цепочках ионов \cite{akerman2015universal} в радиочастотных ловушках Пауля. Лазеры, использующиеся для проведения вычислительных квантовых операций, должны обладать чрезвычайно низкими фазовыми шумами в довольно широкой полосе (порядка 1 МГц), а несущая -- быть как можно более узкой. От степени подавления фазовых шумов такого ``кубитного'' лазера зависит точность квантовых операций \cite{galstyan2025injection}. 
Beyond the aforementioned applications, a new class of tasks has emerged where the laser noise spectrum becomes critically important. These include quantum computing with cold atoms in optical tweezers \cite{levine2018high} or laser-cooled ion chains in Paul RF traps \cite{akerman2015universal}. Lasers used for quantum computational operations must exhibit extremely low phase noise across a relatively broad bandwidth ($\sim$1 MHz) while maintaining an ultra-narrow carrier linewidth. The fidelity of quantum operations depends directly on the phase noise suppression of such "qubit" lasers \cite{galstyan2025injection,semenin2025improved}.
%В случае же необходимости спектроскопии нескольких узких атомных переходов, как это требуется, например, в модели связанных квантовых мемристоров \cite{stremoukhov2024model}, появляется необходимость использования нескольких стабилизированных лазерных источников. Причем в случае реализации квантового мемристора, предложенного в \cite{stremoukhov2024model} от лазерных источников требуется не только низкий уровень фазовых шумов, необходимых для обеспечения достоверности операций, но и метрологические характеристики, аналогичные характеристикам часовых лазерных систем, поскольку в предложенной модели используется два сильно запрещенных перехода в ионе иттербия-171: квадрупольный на длине волны 435 нм с естественной шириной линии 3,1 Гц и октупольный на длине волны 467 нм с естественной шириной линии в единицы нГц.
When spectroscopy of multiple narrow atomic transitions is required, as in the coupled quantum memristor model \cite{stremoukhov2024model}, multiple stabilized laser sources become necessary. For the quantum memristor implementation proposed in \cite{stremoukhov2024model}, the laser sources must not only achieve low phase noise (ensuring operational reliability) but also meet metrological specifications comparable to clock laser systems. This is because the proposed model utilizes two highly forbidden transitions in $^{171}$Yb$^+$ ions: a quadrupole transition at 435 nm (natural linewidth 3.1 Hz) and an octupole transition at 467 nm (natural linewidth in the nHz range).

%Таким образом, при разработке транспортируемых оптических часов и квантовых вычислителей на основе атомов и ионов часто возникает необходимость в использовании набора высокостабильных лазерных генераторов. Одним из возможных решений является создание отдельного опорного резонатора для каждого лазера, однако это приведет к усложнению и увеличению габаритов системы. Альтернативный подход заключается в использовании единого основания-тела для крепления нескольких пар зеркал, образующих опорные резонаторы, рассчитанные на все необходимые в эксперименте длины волн \cite{dawel2024coherent,hill2021dual}. Такой метод позволяет ограничиться одной вакуумной камерой, что упрощает конструкцию, но при этом требует разработки отдельных систем стабилизации для каждого лазера. Перспективным подходом является использование фемтосекундной гребенки оптических частот, распределяющей стабильность частоты одного лазерного источника на различные длины волн. Такой метод является универсальным, упрощает архитектуру системы и требует только одного опорного резонатора.
Therefore, the development of transportable optical clocks and quantum computers based on atoms and ions often requires a set of highly stable laser oscillators. One possible solution involves creating a separate reference cavity for each laser, but this would complicate the system and increase its size. An alternative approach is to use a single monolithic base to mount multiple mirror pairs, forming reference cavities designed for all required wavelengths \cite{dawel2024coherent,hill2021dual}. This method allows the use of a single vacuum chamber,  though it still requires developing individual stabilization systems for each laser.
A promising alternative is the use of a femtosecond optical frequency comb to distribute the frequency stability of a single laser source across multiple wavelengths. This universal approach simplifies the system architecture and requires only one reference cavity.

\begin{figure}
    \centering
    \includegraphics[width=\linewidth]{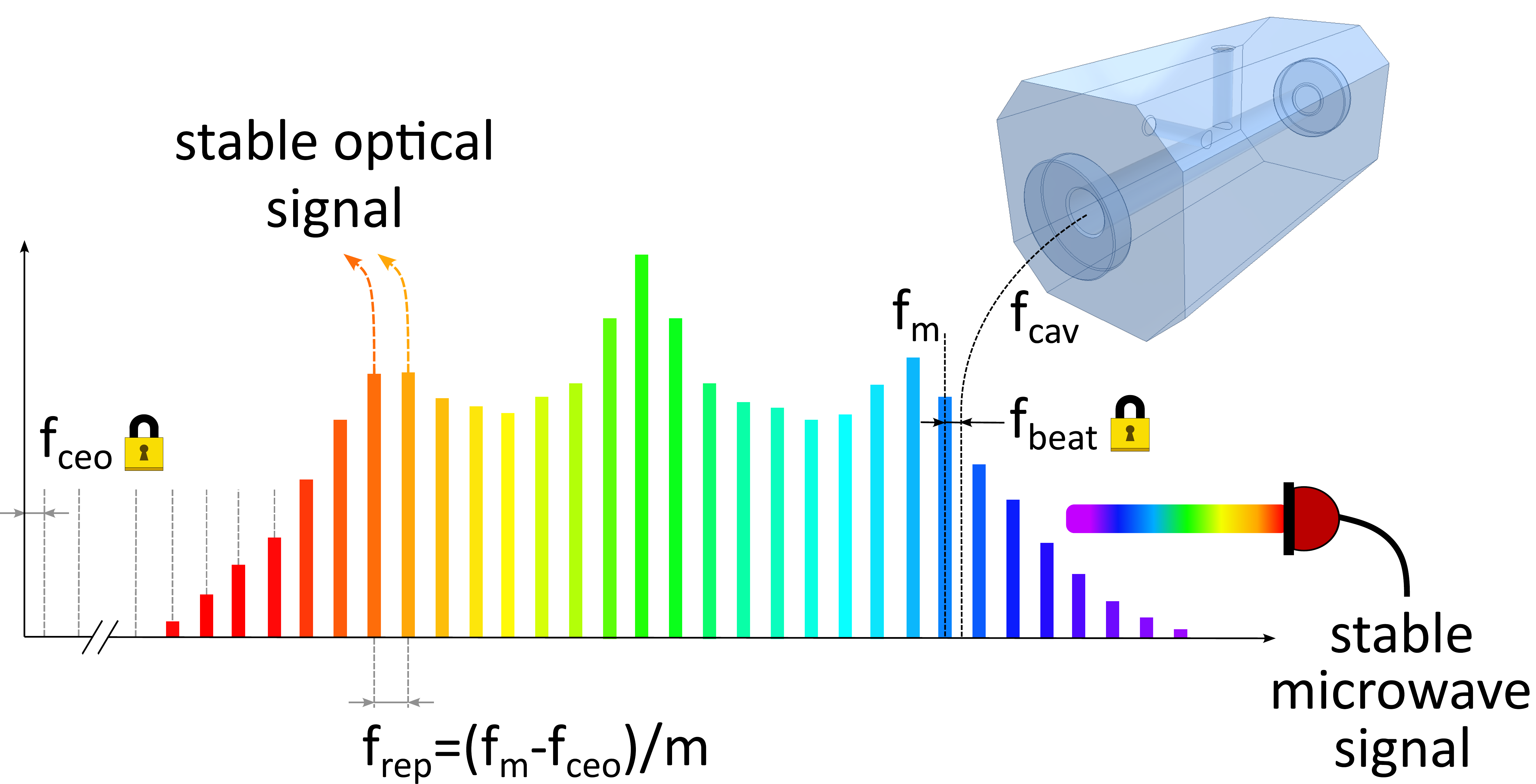}
    \caption{Frequency stability transfer scheme using a femtosecond frequency comb stabilized to an ultrastable reference cavity laser. Each laser frequency is stabilized by phase-locking it to a corresponding comb spectral component via PLLs. Additionally, this scheme enables the synthesis of a highly stable microwave signal by utilizing the comb repetition rate.}
    %Рисунок 1. Схема распределения стабильности частоты при помощифемтосекундной гребенки, стабилизированной по ультрастабильному лазеру с опорным резонатором. Частота каждого лазера стабилизируется путем его привязки к соответствующей спектральной компоненте гребенки через петли ФАПЧ. Кроме того, такая схема позволяет синтезировать высокостабильный микроволновый сигнал, используя частоту повторения импульсов гребенки.}
    \label{fig:Figure_1}
\end{figure}

%Схема, позволяющая перенести характеристики стабильности частоты одного лазера на отличную длину волны в оптическом или микроволновом диапазоне, показана на Рисунке \ref{fig:comb to mw}. 
%Частоты спектральных компонент фемтосекундной оптической гребенки $f_{m}$ определяются частотой повторения импульсов $f_{rep}$ и частотой отстройки $f_{ceo}$.
%Стабилизацию частоты повторений можно осуществить привязкой одной из спектральных компонент гребенки к частоте лазера с опорным резонатором $f_{cav}$. Для этого используются петля фазовой автоподстройки частоты биений (ФАПЧ):
The scheme enabling the transfer of frequency stability characteristics from one laser to a different wavelength in the optical or microwave range is shown in Figure \ref{fig:Figure_1}.
The frequencies of the femtosecond optical frequency comb spectral components $f_{m}$ are determined by the pulse repetition rate $f_{rep}$ and carrier-envelope offset frequency $f_{ceo}$.
Stabilization of the repetition rate can be achieved by phase-locking one of the comb spectral components to the reference cavity-stabilized laser frequency $f_{cav}$. This is implemented using a phase-locked loop (PLL) for beat signal:
\begin{equation}
f_{beat} = f_{cav}-f_{m} = f_{cav}-(f_{ceo}+mf_{rep}).
\end{equation}
%Стандартным подходом для стабилизации частоты отстройки является её измерение при помощи $f-2f$ интерферометра и привязка к частоте повторений, либо к независимому высокостабильному радиочастотному источнику.
%Отдельные спектральные компоненты оптической гребенки могут быть задействованы для передачи стабильности частоты другим лазерам \cite{Benkler2019}. Это достигается за счет организации петли ФАПЧ, которая позволяет синхронизировать частоту целевого лазера с выбранной компонентой гребенки.
%Периодичность следования лазерных импульсов может быть использована и для реализации микроволнового стандарта частоты. Для этого излучение гребенки направляют на быстрый фотодетектор (например, pin-диод), который формирует сигнал на частоте повторений и её гармониках.
The standard approach for carrier-envelope offset (CEO) frequency stabilization involves its measurement using an $f-2f$ interferometer and phase-locking it to either the repetition rate or an independent highly stable RF reference source.
Individual spectral components of the optical frequency comb can be utilized to transfer frequency stability to other lasers \cite{Benkler2019}. This is achieved by implementing a PLL that synchronizes the target laser frequency with a selected comb component.
The periodic pulse train can also be employed to generate a microwave frequency standard. For this purpose, the comb output is directed to a fast photodetector (e.g., a PIN diode), which produces an electrical signal at the repetition rate and its harmonics.

%В работе описывается созданная компактная лазерная система на длине волны 871 нм на основе диодного лазера с внешним резонатором со стабилизацией по высокодобротному интерферометру Фабри-Перо из стекла ULE (Ultra-Low Expansion) и процесс передачи его стабильности при помощи портативной оптической гребенки. Стабильность частоты гребенки охарактеризована методом треуголки \cite{кудеяров2021сличение} при сличении с альтернативными высокостабильными лазерами.
This work describes a compact 871 nm laser system based on an external-cavity diode laser (ECDL) stabilized to a high-finesse Fabry-Pérot cavity made of ultra-low expansion (ULE) glass, along with the process of transferring its stability using a portable optical frequency comb. The comb frequency stability was characterized via the three-cornered hat technique \cite{kudeyarov2021comparison} with other ultrastable lasers.

\section{System design}
\subsection{Laser source}
%В качестве источника лазерного излучения был разработан полупроводниковый лазер с внешним резонатором, реализованным по схеме Литтрова \cite{de1993mode}. Длина внешнего резонатора \(L\) - важная величина, во многом определяющая характеристики лазерного излучения. Ширина спектральной линии излучения такого лазера определяется выражением
As the laser source, we developed an external cavity diode laser (ECDL) in a Littrow configuration \cite{de1993mode}. The external cavity length \(L\)  is a critical parameter that largely determines the laser output characteristics. The spectral linewidth of such a laser is given by:

\begin{equation}
\Delta f = \frac{\Delta f_{LD}}{(1+\frac{L}{nL_{LD}})^2},
\end{equation}
%где \(\Delta f_{LD}\) - спектральная ширина линии лазерного диода, а \(n\) и \(L_{LD}\) - его показатель преломления и длина, соответственно \cite{schawlow2002infrared}. Таким образом, за счёт удлинения резонатора можно заметно сузить ширину линии лазера и подавить его фазовые шумы \cite{kolachevsky2011low}, однако это зачастую сопряжено с увеличением чувствительности к акустическим колебаниям и как следствие, к увеличению фазовых шумов и скачкам моды. В ряде работ длина внешнего резонатора ограничивается величиной 20 - 30 мм \cite{vassiliev2006compact,dutta2012mode,chen2017single,dutta2023stable}. 
where \(\Delta f_{LD}\) is the laser diode linewidth, and \(n\) and \(L_{LD}\) represent its refractive index and length, respectively \cite{schawlow2002infrared}. Thus, extending the cavity can significantly narrow the laser linewidth and suppress phase noise \cite{kolachevsky2011low}. However, this typically increases susceptibility to acoustic vibrations, consequently raising phase noise and inducing mode hops. In several studies, the external cavity length is limited to 20–30 mm \cite{vassiliev2006compact,dutta2012mode,chen2017single,dutta2023stable}.

%Другим важным параметром лазера с внешним резонатором является диапазон непрерывной перестройки частоты без скачков моды. Обеспечить достаточно большую длину резонатора одновременно с высокой механической стабильностью конструкции и широким диапазоном перестройки позволяет дизайн, описанный в \cite{Heine2011,kirilov2015compact}, и примененный в данной работе.
Another critical parameter of external-cavity lasers is the continuous frequency tuning range without mode hops. The design implemented in this work enables both sufficiently long resonator length and wide tuning range while maintaining high mechanical stability \cite{Heine2011,kirilov2015compact}.

\begin{figure}
    \centering
    \includegraphics[width=0.9\linewidth]{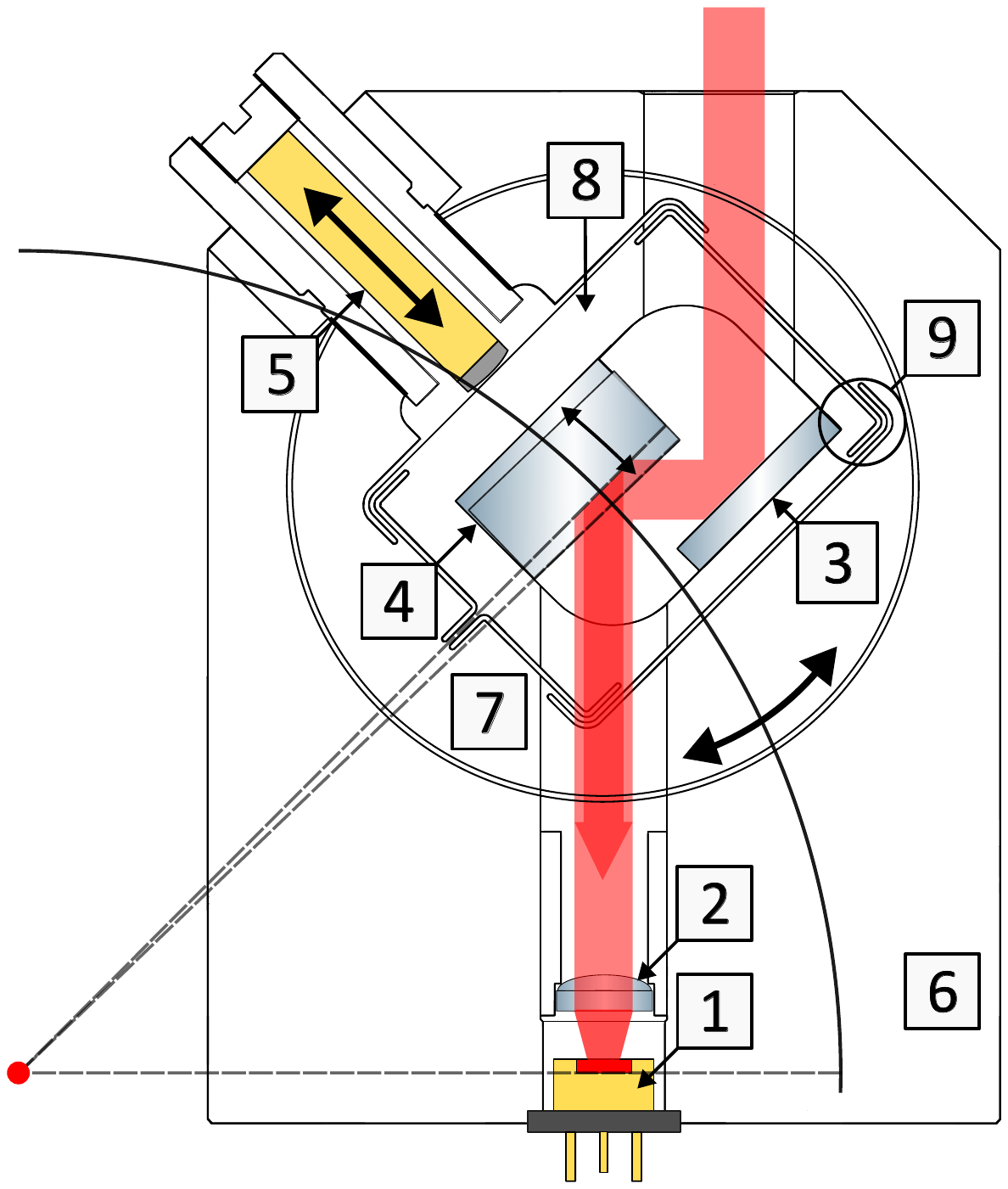}
    \caption{
    %Рисунок 2. Принципиальное устройство лазера схемы Литтрова с вынесенной осью вращения дифракционной решетки.  1 - лазерный диод, 2 - асферическая линза, 3 - зеркало, 4 - дифракционная решётка 1200 штрихов/мм, 5 - пьезоактюатор, 6 - корпус лазера, 7 - база для крепления решётки, 8 - подвижная часть базы, 9 - перемычка. На данной схеме не показаны части лазера, отвечающие за юстировку лазерного диода, линзы и поворот решетки в широком диапазоне.
    Schematic of a Littrow configuration laser with external diffraction grating rotation axis: 1 - laser diode, 2 - aspheric lens, 3 - mirror, 4 - diffraction grating, 5 - piezoelectric actuator, 6 - laser housing, 7 - grating mounting base, 8 - movable frame, 9 - flexure joint. This schematic omits components for laser diode alignment, lens adjustment, and coarse grating rotation.}
    \label{fig:Figure_2}
\end{figure}

%Принципиальная схема лазера представлена на Рисунке \ref{fig:laser}. Внешний резонатор формируется между просветленным лазерным диодом, излучение которого коллимируется асферической линзой с фокусным расстоянием 4.6 мм, и дифракционной решёткой с плотностью штрихов 1200 мм$^{-1}$, отражающей свет в -1 порядок. Чтобы обеспечить отсутствие скачков моды, при регулировке длины волны необходимо согласовывать изменение длины резонатора и угол поворота дифракционной решётки \cite{de1993mode,saliba2009mode,hult2005wide}. Для этого решётка должна совершать круговое движение относительно оси вращения, лежащей на пересечении плоскостей диода и решётки. Для осуществления такого типа движения подвижная часть базы, на которой закреплена дифракционная решётка, соединена с основанием тонкими упругими перемычками, задающими траекторию вращения. Размеры и положение перемычек были оптимизированы в процессе моделирования методом конечных элементов для обеспечения верной траектории движения решетки. Поворот решетки может выполняться с помощью микрометрического винта, вращающего базу с закрепленной решеткой, для грубой подстройки и пьезоактюатором для тонкой подстройки. Все составные части лазера закреплены на монолитном алюминиевом корпусе.
The laser schematic is shown in Figure \ref{fig:Figure_2}. The external cavity is formed between an anti-reflection coated laser diode (collimated using a 4.6 mm focal length aspheric lens) and a 1200 lines/mm diffraction grating reflecting light in the -1st order. To prevent mode hops during wavelength tuning, the cavity length adjustment must be synchronized with the grating rotation angle \cite{de1993mode,saliba2009mode,hult2005wide}. This requires the grating to rotate around an axis lying at the intersection of the diode and grating planes.
The movable frame to which the grating is attached is connected to the base via thin flexure joints that constrain its motion to the required circular trajectory. The joint dimensions and positions were optimized through finite element modeling to ensure proper grating movement. Coarse wavelength tuning is achieved via a micrometer screw pushing the frame, while fine adjustment uses a piezoelectric actuator. All components are mounted in a monolithic aluminum housing.

%Реализация схемы крепления дифракционной решётки с выносом оси вращения за пределы корпуса лазера позволила сделать его компактным: внешний резонатор имеет длину около 40 мм, при этом габариты лазера составляют 90$\times$100$\times$75 мм. Диапазон непрерывной перестройки частоты при помощи пьезоактюатора составил более 6 ГГц. 
The implementation of a diffraction grating mounting scheme with an external rotation axis enabled a compact design: the cavity maintains a 40 mm length while the overall laser dimensions are just 90$\times$100$\times$75 mm. The piezoelectric actuator provided a continuous frequency tuning range exceeding 6 GHz.
%Помимо прочего, такая структура обладает высокой устойчивостью к механическим и акустическим колебаниям. Это достигается за счёт симметрично расположенных упругих соединений базы решётки с её подвижной частью и малых размеров лазера.
The design demonstrates comparatively low sensitivity to vibrations due to compactness and structure rigidity.

%Система температурной стабилизации лазера состоит из двух частей: пассивной, состоящей из алюминиевого корпуса, и активной, состоящей из элемента Пельтье и терморезистора. Алюминиевый корпус позволяет изолировать лазерный диод от колебаний температуры окружающей среды. Активная часть дает возможность установить заданную температуру лазера с помощью ПИД-регулятора. Также в лазер был интегрирован электронный модуль, позволяющий реализовать быструю петлю обратной связи с помощью шунтирования тока инжекции внешним сигналом.
The laser thermal stabilization system comprises two parts: a passive (aluminum housing) and an active (Peltier element with thermistor). The aluminum housing isolates the laser diode from external temperature fluctuations, while the active system enables precise temperature control via a PID controller. An integrated electronic module also provides fast laser frequency feedback by shunting the injection current with an external control signal.

%Длина волны излучения лазера была выбрана равной 871 нм, что позволяет использовать вторую гармонику его излучения для возбуждения часового перехода в ионе $^{171}$Yb$^+$. Мощность излучения при токе инжекции 90 мА составила 13 мВт. Ширина линии излучения без дополнительной стабилизации была определена по сигналу биений с опорным ультрастабильным лазером на длине волны 871 нм \cite{Zalivako2020} и составила 840 кГц при времени измерения 1 мс.
The laser was designed to work at 871 nm to allow the use of its second harmonic for exciting the clock transition in $^{171}$Yb$^+$ ions. At an injection current of 90 mA, the output power reached 13 mW. The emission linewidth without active stabilization was measured by beating the laser against an 871 nm ultrastable reference laser \cite{Zalivako2020}, yielding 840 kHz at 1 ms measurement time.

\subsection{Reference cavity}
%В качестве опорного резонатора используется интерферометр Фабри-Перо из ULE-стекла длиной 100 мм с горизонтальной конфигурацией подвеса. При условии оптимизации системы подвеса горизонтальная ориентация позволяет снизить восприимчивость резонатора к ускорениям \cite{жаднов2018пределе}, что исключительно важно для возимых систем. 
The reference cavity is a 100 mm long Fabry-Pérot interferometer made of ULE glass, configured in a horizontal mounting orientation. With an optimized suspension design, this horizontal configuration reduces the sensitivity of the cavity to acceleration \cite{zhadnov2018thermal}. This is a critical feature for transportable systems.

%Зеркала резонатора сформированы 36 слоями Ta\textsubscript{2}O\textsubscript{5}/SiO\textsubscript{2}, напылёнными на подложки из ULE-стекла. Для устойчивости резонатора и удобства завода излучения используется пара из плоского и сферического зеркал. Сферическое зеркало имеет радиус кривизны 1 м. Резкость резонатора составляет около 180 000. Девиация Аллана относительных флуктуаций частоты моды резонатора, вызванных тепловыми шумами, приблизительно равна $7\times 10^{-16}\ $ и определяется колебаниями поверхности ULE-подложек зеркал.
The cavity mirrors consist of 36 alternating Ta\textsubscript{2}O\textsubscript{5}/SiO\textsubscript{2} layers deposited on ULE substrates. A plano-spherical mirror configuration (spherical mirror radius of curvature is 1 m) ensures resonator stability and simplifies optical coupling. The cavity exhibits a finesse of $\sim$180 000. The Allan deviation of the mode fractional frequency fluctuations due to thermal noise is $\sim7\times 10^{-16}$, primarily limited by ULE substrate surface fluctuations.

\begin{figure}
    \centering
    \includegraphics[width=\linewidth]{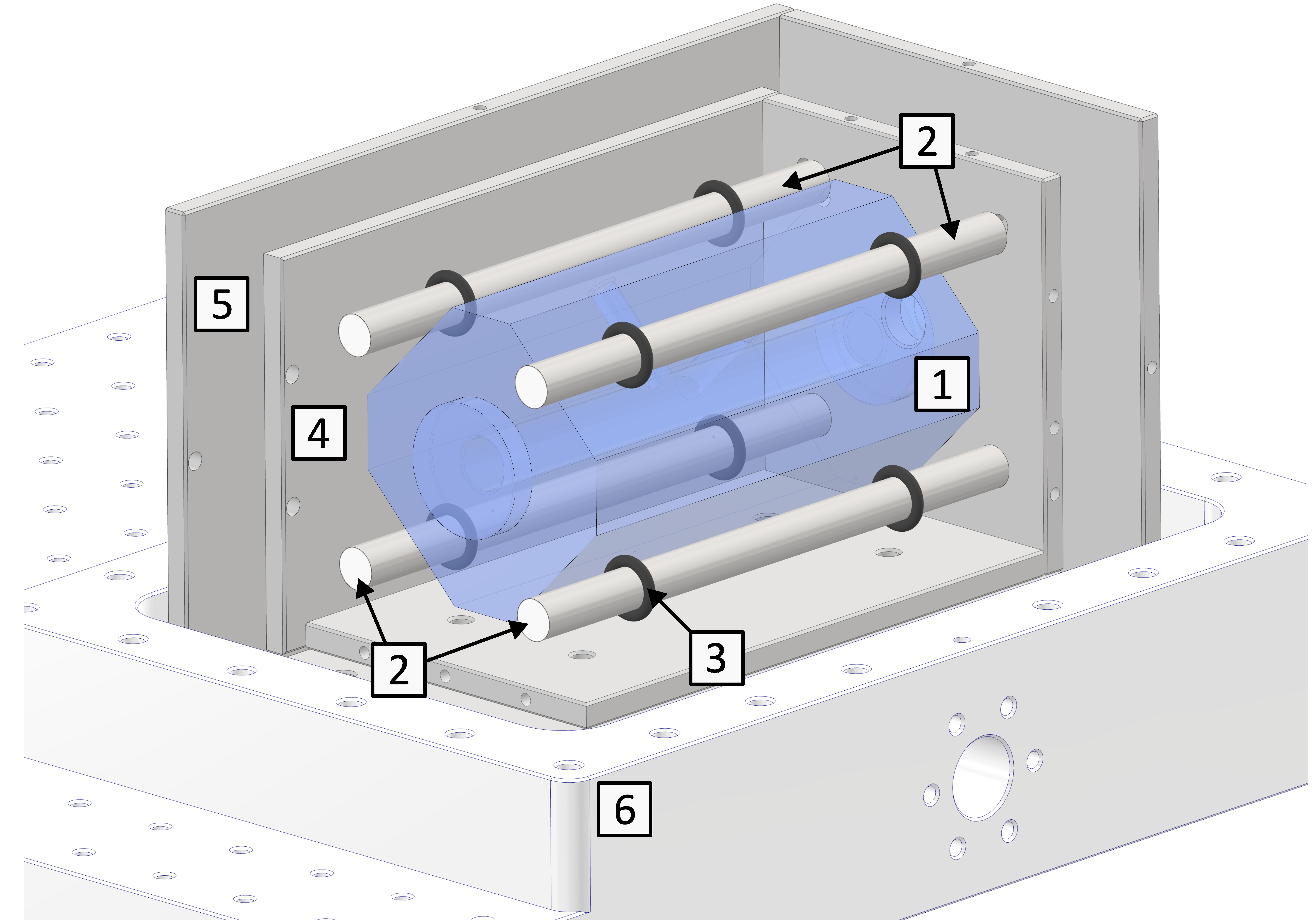}
    \caption{
    %Рисунок 3. Вид вакуумной камеры резонатора в разрезе . 1 - резонатор Фабри-Перо, 2, - направляющие, 3 - кольца из вакуумной резины (витон), 4 - пассивный тепловой экран, 5 - активный тепловой экран, 6 - основание вакуумной камеры.
Cross-sectional view of the reference cavity vacuum chamber: 1 – Fabry-Pérot cavity, 2 – guide rods, 3 – vacuum-grade rubber O-rings (Viton), 4 – passive thermal shield, 5 – active thermal shield, 6 – vacuum chamber base.}
    \label{fig:Figure_3}
\end{figure}

%Подвес резонатора представляет собой систему из четырех направляющих, расположенных вдоль оси резонатора и закрепленных на торцевых стенках внутреннего теплового экран (Рисунок \ref{fig:chamber}). Тело резонатора фиксируется при помощи восьми витоновых колец, надетых на направляющие. Положение колец на направляющих оптимизировалось при помощи моделирования для минимизации вибрационной чувствительности. Система подвеса проектировалась таким образом, чтобы первый механический резонанс лежал значительно выше 1 кГц, так как низкочастотные колебания вносят наибольший вклад в нестабильность длины резонатора на требуемых для большинства применений временах усреднения \cite{keller2014simple}. 
The cavity suspension system comprises four guide rods aligned along the cavity axis and mounted to the end walls of the inner thermal shield (Figure \ref{fig:Figure_3}). The cavity body is fixed using eight Viton O-rings fitted over the rods. The O-ring positions were optimized through finite-element modeling to minimize vibrational sensitivity. The suspension was designed to place the first mechanical resonance well above 1 kHz, as low-frequency vibrations dominate cavity length instability at the averaging times relevant for most applications \cite{keller2014simple}.

%Вакуумная камера резонатора изготовлена из дюралюминия с индиевыми уплотнениями. Данная конфигурация позволяет поддерживать вакуум на уровне $5\times10^{-9}$ мБар при помощи двух ионных насосов с производительностью 3 л/c. Температурная стабилизация резонатора осуществлена с помощью двух тепловых экранов: температура внешнего стабилизируется активно при помощи элемента Пельтье и терморезистора, а внутренний выполняет функцию пассивной стабилизации. 
The cavity vacuum chamber is constructed from duralumin with indium seals, maintaining a vacuum level of $5\times10^{-9}$ mBar using two ion pumps (3 l/s each). Thermal stabilization employs two shields: the outer shield temperature is actively controlled via a Peltier element and thermistor, while the inner shield provides passive stabilization.

\subsection{Frequency stabilization system}
%Для повышения вибрационной устойчивости и компактности лазерной системы была выбрана компоновка оптической схемы и опорного резонатора на единой базе, которая является основанием вакуумной камеры резонатора. Такое техническое решение позволило разместить всю ультрастабильную лазерную систему в габаритах 300$\times$300$\times$200 мм. Основные характеристики системы приведены в Таблице \ref{tab:char}.
To enhance vibration sustainability and compactness, the optical layout and reference cavity were integrated onto a single baseplate serving as the cavity vacuum chamber foundation.
This design enabled full integration of the ultrastable laser system within dimensions of 300$\times$300$\times$200 mm (18 L), with a total weight of just 15 kg.

%Стабилизация лазерного излучения осуществляется методом ПДХ. Для подавления паразитной оптической обратной связи между элементами системы и  лазерным источником схема содержит  оптический изолятор и акустооптический модулятор, сдвигающий частоту излучения на 80 МГц. Фазовая модуляция излучения на частоте около 20 МГц осуществляется при помощи электрооптического модулятора (ЭОМ) с гранями под углом Брюстера, что позволяет подавить влияние остаточной амплитудной модуляции \cite{Bi2019}. Стабилизация частоты осуществляется при помощи быстрой и медленной петель обратной связи, управляющих током лазера и напряжением на его пьезоактюаторе, соответственно. Ширина спектра стабилизированного излучения составила 9 Гц при времени усреднения 1 с.
Laser frequency stabilization is achieved via the PDH method. To suppress parasitic optical feedback between system components and the laser source, the setup incorporates an optical isolator and an acousto-optic modulator (AOM) introducing an 80 MHz frequency shift. Phase modulation at ~20 MHz is implemented using a Brewster cut electro-optic modulator (EOM) to mitigate residual amplitude modulation effects \cite{Bi2019}. Frequency stabilization employs fast and slow feedback loops controlling the laser current and piezoelectric actuator voltage, respectively. The stabilized emission spectrum exhibited a 9 Hz linewidth at 1 s averaging time.

%Важной характеристикой возимой высокостабильной лазерной системы является чувствительность частоты стабилизированного излучения к внешним вибрациям. Для исследования вибрационной чувствительности одновременно записывалась частота биений стабилизированного излучения с опорным ультрастабильным лазером $f$ и показания трехосевого сейсмоприёмника «Геоакустика» А0531 $a_x,a_y,a_z$ (ось $x$ направлена вдоль оптической оси резонатора, ось $y$ перпендикулярно оси резонатора, но параллельно основанию системы, ось $z$ направлена перпендикулярно основанию; см. Рисунок \ref{fig:chamber}). Сейсмоприемник и лазерная система были расположены на оптическом столе на пневматических опорах. Опорный лазер находился в удалённом помещении, и его излучение доставлялось по оптоволоконной линии, что позволило исключить влияние вибраций на частоту его излучения.
%Возмущения вносились качанием оптического стола с частотой приблизительно 1 Гц. Было записано три набора данных для различных доминирующих направлений  возмущений. Все наборы данных совместно аппроксимировались линейной функцией: 
A critical parameter for transportable ultrastable laser systems is their vibration sensitivity. To characterize this, we simultaneously recorded the beat note frequency $\nu$ between the stabilized laser and a reference ultrastable laser and acceleration data $a_x,a_y,a_z$ from a triaxial seismometer (Geoacoustics A0531). Here $x$ is axis along the cavity optical axis,  $y$ - axis perpendicular to cavity axis but parallel to base,  $z$ - axis perpendicular to base (see Figure \ref{fig:Figure_3}). Both systems were mounted on an optical table with pneumatic isolation. The reference laser, stationed in a remote location, delivered light via fiber to eliminate vibration-induced frequency shifts.
Controlled perturbations (at $\sim$1 Hz) were introduced by tilting the optical table. Three datasets were acquired for different dominant perturbation directions, then jointly fitted with a linear model:
\begin{equation}
    \nu = K_x a_x + K_y a_y + K_z a_z + \nu_0.
    \label{eq:force_components}
\end{equation} 

%Полученные коэффициенты чувствительности системы к ускорению представлены в таблице \ref{tab:vibr}. Для возмущений низкой частоты, лежащих ниже первого акустического резонанса, можно пренебречь частотной зависимостью этих коэффициентов. Чувствительность лазерной системы к ускорениям вдоль осей $x$ и $y$ сравнима с показателями транспортируемых систем, использующих резонаторы аналогичного размера \cite{Chen2014, Swierad2016, Hafner2020, Cole2023}. Достаточно высокая чувствительность по оси $z$ может быть связана с недостаточно точной установкой положения витоновых колец при сборке; для снижения чувствительности необходимо провести оптимизацию положения одновременно с измерением чувствительности. Тем не менее, даже текущие характеристики позволяют ожидать от системы нестабильности частоты менее $10^{-14}$ при ускорениях на уровне нескольких $\mu$g, наблюдаемых, например, при бортовом использовании \cite{LI2022112031}.
The measured acceleration sensitivity coefficients are presented in Table \ref{tab:vibr}. For low-frequency perturbations below the first acoustic resonance, these coefficients can be treated as frequency independent. The laser system acceleration sensitivity along the $x$ and $y$ axes is comparable to transportable systems using similarly sized resonators \cite{Chen2014, Swierad2016, Hafner2020, Cole2023}. The relatively high  $z$ axis sensitivity may originate from imperfect Viton O-ring positioning during assembly; sensitivity reduction requires position optimization during real-time sensitivity measurements. Nevertheless, even current performance suggests frequency instability below $10^{-14}$ under typical onboard $\mu$g level accelerations \cite{LI2022112031}.

\begin{table}
\centering
\caption{
%Таблица 2. Коэффициенты чувствительности частоты стабилизированного лазерного излучения к ускорениям, 
Sensitivity coefficients of the stabilized laser radiation frequency to accelerations, $1/g$.} 
\vspace{6mm}
\begin{tabular}{| c | c | c |}
\hline
$K_x$ & $K_y$& $K_z$\\
\hline
$4.1  \times 10^{-10}$& $1.8  \times 10^{-10}$& $2.4  \times 10^{-9}$\\
\hline

\end{tabular}
\label{tab:vibr}
\end{table}

\section{Stability dissemination via femtosecond comb}
%Распределение стабильности через фемтосекундную гребенку

%Для передачи стабильности созданной системы в другой частотный диапазон была использована компактная фемтосекундная гребенка производства ООО «Авеста-Проект». Данная гребенка включает в себя фемтосекундный осциллятор, $f-2f$ интерферометр, генератор суперконтинуума для генерации спектра на длине волны 871 нм и встроенный блок оптических биений для этой длины волны. При подаче на гребенку стабильного оптического сигнала на длине волны 871 нм формируется сигнал биений с ближайшей частотной компонентой гребенки, позволяющий стабилизировать частоту повторения относительно частоты излучения лазера с помощью системы ФАПЧ. Частота смещения гребёнки детектируется с помощью $f-2f$ интерферометра и стабилизируется относительно субгармоники частоты повторения с помощью еще одной независимой системы ФАПЧ. Затем можно стабилизировать частоту другого лазерного источника относительно соответствующей компоненты гребенки. Также гребенка позволяет генерировать высокостабильный микроволновый сигнал на частоте 1 ГГц. Модуль формирования микроволнового сигнала выполнен на основе высокочастотного балансного фотоприемника с последующей фильтрацией гармоник частоты повторения фемтосекундного лазера (125 МГц). Для предотвращения насыщения фотоприемника излучение фемтосекундного лазера пропускается через три последовательных интерферометра Маха-Цандера. Полный объем компактной гребенки составляет 21,7 литра.
To transfer the system frequency stability to other spectral ranges, we employed a compact femtosecond frequency comb made by Avesta Project Ltd. This comb system includes a femtosecond oscillator, an $f-2f$ interferometer, a supercontinuum generator for 871 nm spectral coverage and a built-in 871 nm beat-note detection module. When a stable 871 nm optical signal is injected, the comb generates a beat signal with its nearest spectral component, enabling repetition rate stabilization against the laser frequency via a PLL. The carrier-envelope offset frequency is detected by the $f-2f$ interferometer and stabilized relative to a subharmonic of the repetition frequency via an additional independent PLL system. This allows locking additional laser sources to other comb modes.
The comb also generates a 1 GHz ultrastable microwave signal using a high-frequency balanced photodetector module, which filters the harmonics of the femtosecond laser repetition rate (125 MHz). To prevent photodetector saturation, the femtosecond laser radiation is passed through three cascaded Mach-Zehnder interferometers. The entire comb system occupies a volume of 21.7 l.

%Для исследования достигаемых показателей стабильности была осуществлена передача стабильности волоконному DFB лазеру Koheras BASIK на длине волны 1550 нм. Для подстройки частоты биений в рабочий диапазон излучение лазерной системы 871 нм подавалось на гребенку через волоконный АОМ с номинальной частотой 200 МГц.
To evaluate the achievable stability metrics, frequency stability was transferred to a Koheras BASIK DFB fiber laser at 1550 nm. For beatnote frequency tuning into the operational range, the 871 nm laser output was directed to the frequency comb through a fiber-coupled AOM with a nominal drive frequency of 200 MHz.

\begin{figure}
    \centering
    \includegraphics[width=1\linewidth]{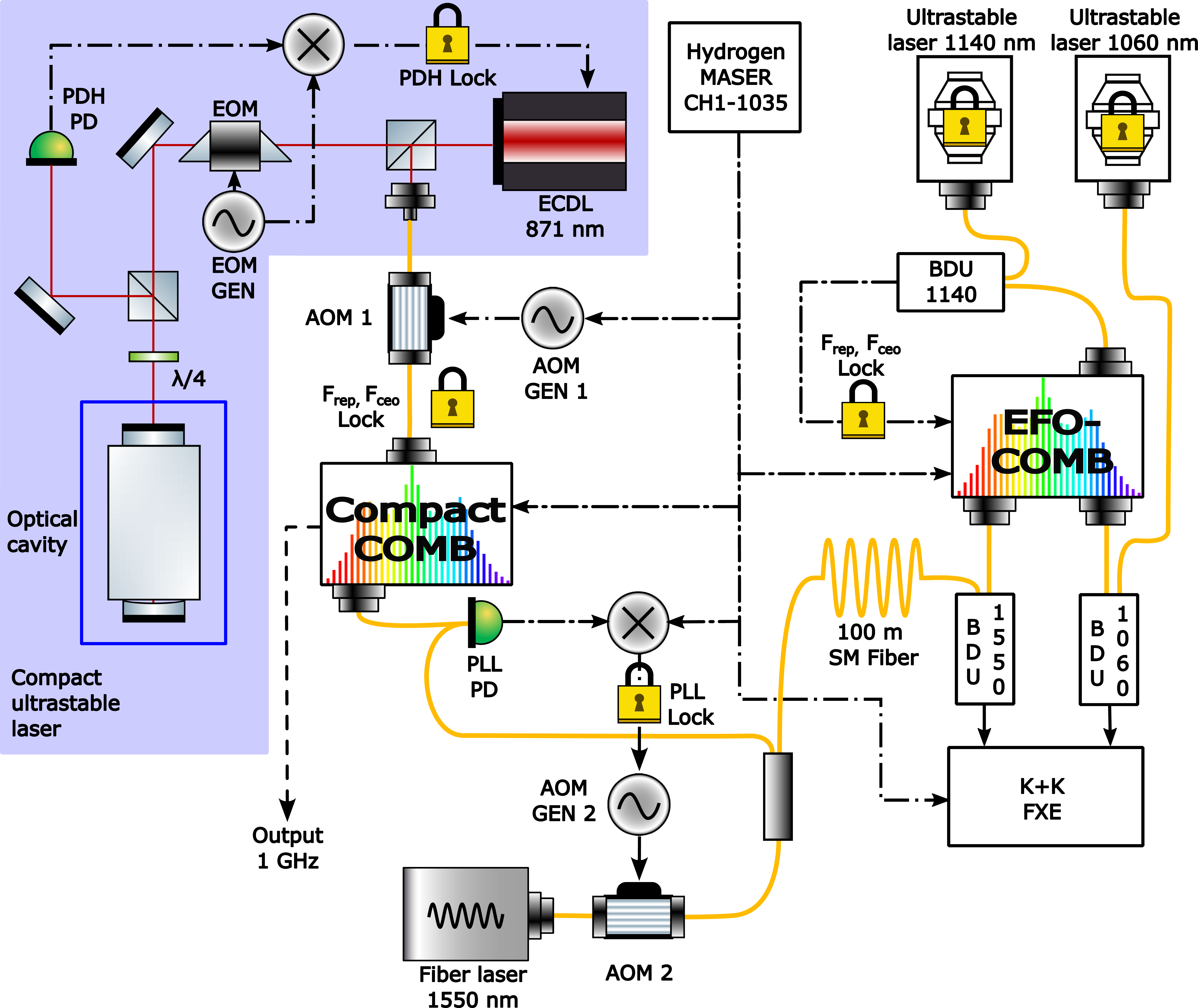}
    \caption{Measurement setup schematic for system stability characterization: ECDL - external-cavity diode laser, AOM - acousto-optic modulator, EOM - electro-optic modulator, $\lambda/4$ - quarter-wave plate, GEN - RF signal generator, PLL - phase-locked loop, PD - photodetector, BDU - beat detection unit.}
    %Рисунок 4. Общая схема измерения стабильности системы. ECDL - полупроводниковый лазер с внешним резонатором, AOM - акустооптический модулятор, EOM - электрооптический модулятор,  $\lambda/4$ - четвертьволновая фазовая пластинка, GEN - радиочастотный генератор, PLL - ФАПЧ, PD - фотодетектор, BDU - блок оптических биений. 

%Figure 4. Measurement setup schematic for system stability characterization: ECDL - external-cavity diode laser, AOM - acousto-optic modulator, EOM - electro-optic modulator, λ/4 - quarter-wave plate, GEN - RF signal generator, PLL - phase-locked loop, PD - photodetector, BDU - beat detection unit.
    \label{fig:Figure_4}
\end{figure}

%Стабильность частоты волоконного лазера измерялась методом «треуголки» \cite{Allan} относительно частот двух опорных ультрастабильных лазеров на длинах волн 1140 нм и 1060 нм  с использованием лабораторной фемтосекундной гребенки Avesta EFO-COMB. Конструкция обеих опорных лазерных систем аналогична приведенной в \cite{golovizin2019ultrastable}. Схема измерения представлена на Рисунке \ref{fig:scheme}. Излучение лазера 1550 нм передавалось в лабораторию, в которой были расположены опорные лазеры и лабораторная гребенка, по одномодовой волоконной линии длиной 100 метров.  На другом конце линии формировался сигнал биений этого излучения с излучением гребенки EFO-COMB, стабилизированной по сигналу лазера 1140 нм. Частота биений записывалась измерителем фазы K+K FXE с высоким разрешением без мертвого времени \cite{kramer2001multi}. Аналогичным образом одновременно измерялась частота биений лазера на длине волны 1060 нм с гребенкой. Все биения формировались при помощи блоков оптических биений на основе балансных фотодетекторов. В качестве радиочастотной опоры для всех систем ФАПЧ, генераторов для питания АОМ и измерителя фазы использовался сигнал на частоте 10 МГц активного водородного мазера.
The fiber laser frequency stability was characterized via a three-cornered hat measurement \cite{Allan} against two reference ultrastable lasers (1140 nm and 1060 nm) using an Avesta EFO-COMB femtosecond comb. Both reference lasers employed designs identical to \cite{golovizin2019ultrastable}. The measurement setup is shown in Figure \ref{fig:Figure_4}. 
The 1550 nm laser light was delivered to the laboratory with the reference lasers and comb via a 100-meter single-mode fiber link. At the far end of the link, a beat signal was generated between this radiation and the output of the EFO-COMB, stabilized to an 1140 nm laser signal. The beat frequency was recorded with high resolution and no dead time using a K+K FXE frequency counter \cite{kramer2001multi}. Simultaneously, the beat frequency between a 1060 nm wavelength laser and the frequency comb was measured in the same manner. All beat signals were generated using optical beat units based on balanced photodetectors. A 10 MHz signal from an active hydrogen maser served as the RF reference for all phase-locked loops, AOM drive generators, and the frequency counter.

\begin{figure}
    \centering
    \includegraphics[width=\linewidth]{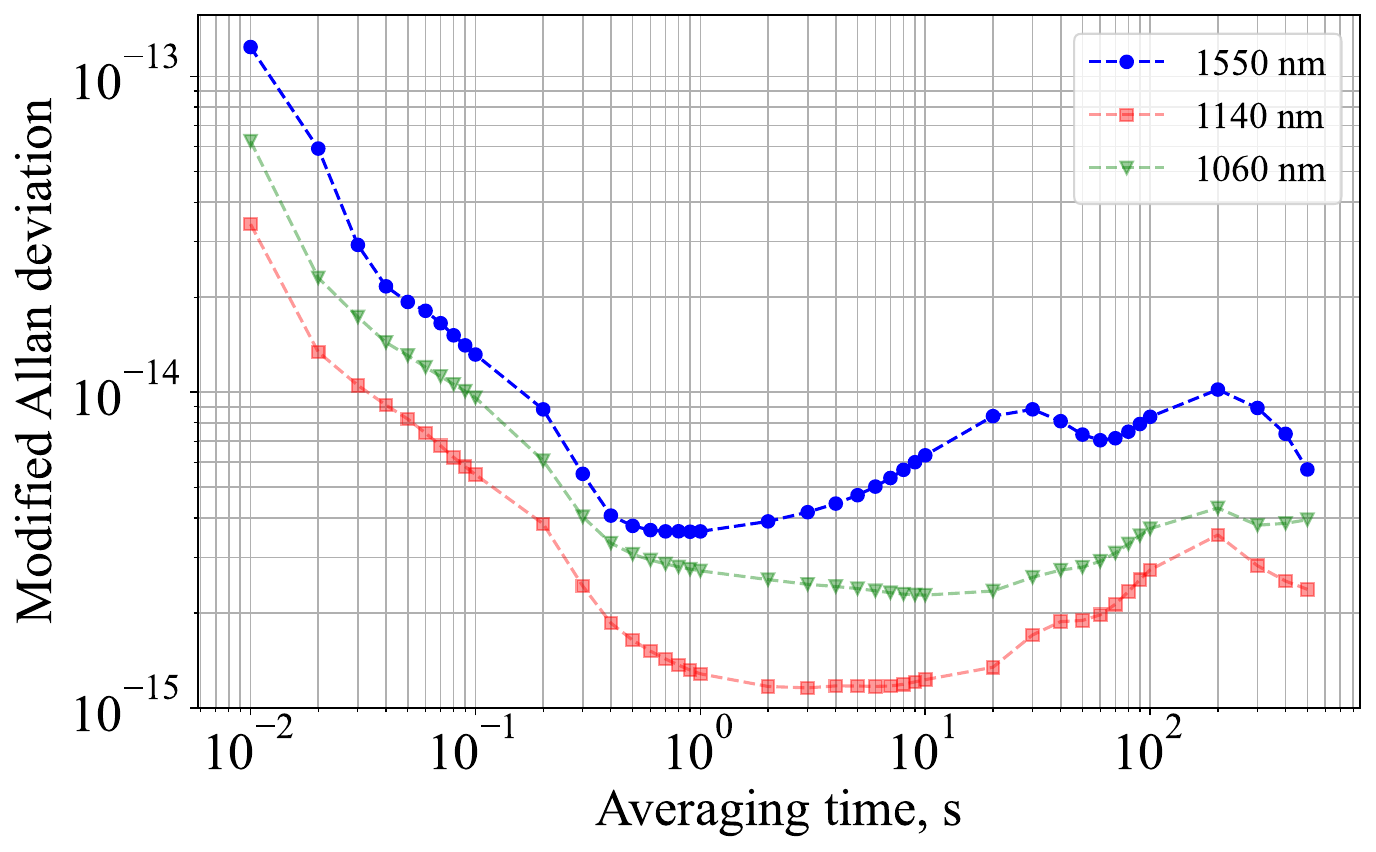}
    \caption{
    %Рисунок 5.  Относительная нестабильность частоты лазера на длине волны 1550 нм, стабилизированного по опорному лазеру на длине волны 871 нм через фемтосекундную гребенку (синий), а также использованных для метода «треуголки» ультрастабильных лазеров на длинах волн 1060 нм (зеленый) и 1140 нм (красный). Данные приведены с учетом вычета линейного дрейфа частоты.
    Fractional frequency instability of the 1550 nm laser (blue) stabilized to the 871 nm reference laser using a femtosecond frequency comb, reference ultrastable lasers at 1060 nm (green) and 1140 nm (red) used for the three-cornered hat method. All data accounts for linear frequency drift removal.}
   
    \label{fig:Figure_5}
\end{figure}

%После вычета линейного дрейфа (для биений гребенки и исследуемой системы он составил 195 мГц/c) была рассчитана модифицированная девиация Аллана для каждого из лазеров (Рисунок \ref{fig:dev}). Относительная нестабильность частоты исследуемой системы составила менее $4\times10^{-15}$ на временах усреднения 0.4 - 2 с и менее $10^{-14}$ на временах усреднения 0.2 - 500 с. Данные величины можно считать консервативной оценкой нестабильности, переданной лазеру 1550 нм от ультрастабильной лазерной системы 871 нм, так как относительная нестабильность, вносимая в сигнал ФАПЧ и оптоволоконной линией, составляет менее $10^{-15}$ в обсуждаемом временном диапазоне.  
After removing the linear drift (195 mHz/s for a beat between the frequency comb and the system under test), we calculated the modified Allan deviation for each laser (Figure \ref{fig:Figure_5}). The fractional frequency instability of the test system was below $4\times10^{-15}$ for averaging times of 0.4–2 s and under $10^{-14}$ for 0.2–500~s. These values represent a conservative estimate of the instability transferred from the 871 nm ultrastable laser system to the 1550 nm laser, as the fractional instability contributed by the PLL and fiber link noise remains below $10^{-15}$ across the studied time range.

\section{Conclusion}

%В рамках данной работы разработана и успешно продемонстрирована компактная лазерная система, обеспечивающая передачу стабильности частоты от лазера с длиной волны излучения 871 нм к лазеру с длиной волны излучения 1550 нм. При помощи небольших модификаций фемтосекундного синтезатора стабильность частоты может быть передана и на другие длины волн видимого и инфракрасного диапазона. Система состоит из двух модулей: ультрастабильного лазера, стабилизированного по моде вакуумного резонатора Фабри-Перо, и фемтосекундной гребенки, которая стабилизируется по этому лазеру. Модули имеют объем менее $18\ $и $25\ $л, соответственно. Их компактность обеспечивается плотным расположением оптических, вакуумных, электронных компонентов и лазеров на единой оптической плите. Измеренная вибрационная чувствительность частоты моды резонатора по различным осям находится на уровне $10^{-9}/g $ и ниже. Девиация Аллана относительных флуктуаций частоты моды стабилизированной фемтосекундной гребенки при сличении с двумя альтернативными ультрастабильными лазерами достигает $3.6\times10^{-15}$ за время усреднения 1 с и составляет менее $10^{-14}$ на временах усреднения 0.2 - 500 с.

This work demonstrates a compact laser system that successfully transfers frequency stability from an 871 nm laser to a 1550 nm laser. With minor modifications to the femtosecond frequency comb, this stability can be extended to other visible and infrared wavelengths. The system comprises two modules: an ultrastable laser locked to a Fabry-Pérot reference cavity and a femtosecond comb stabilized to this laser. With volumes under 18 l and 22 l respectively, their compactness results from dense integration of optical, vacuum, electronic, and laser components on a single optical bench. The measured cavity mode vibration sensitivity is below $3\times10^{-9}/g $ across all axes. When compared against two independent ultrastable lasers, the comb exhibits a modified Allan deviation of $3.6\times10^{-15}$ at 1 s averaging time and $<10^{-14}$ for 0.2–500 s.

%Собранная и охарактеризованная в ФИАН система была перевезена в другую лабораторию, находящуюся на расстоянии 15 км на автомобиле, где подтвердила свою работоспособность. 
After characterization at Lebedev Physical Institute, the system was transported to another laboratory 15 km away and maintained full functionality.
%Продемонстрированный в работе принцип позволяет существенно упростить устройство и уменьшить габариты оптических часов и квантовых компьютеров на атомах и ионах, сократив необходимое количество стабилизированных лазерных систем до одной.
%Описанный в работе прибор может быть использован в качестве элемента оптических часов на ионе иттербия, лазера для выполнения квантовых операций с ионами в квантовом компьютере, а также для генерации высокостабильного микроволнового сигнала. 
The demonstrated approach significantly simplifies the architecture and reduces the size of optical atomic clocks and ion/atom-based quantum computers by minimizing the required number of stabilized laser systems to just one.
The developed device can serve as a core component of ytterbium ion optical clocks, a laser for operations in ion-trap quantum computers, or a high-stability microwave signal generator.

\ack
The study was supported by the Russian Science Foundation grant No. 24-12-00415, https://rscf.ru/project/ 24-12-00415/. 

%\textbf{Конфликт интересов.}
%Авторы данной работы заявляют, что у них нет конфликта интересов.
 \section*{References}
 
\bibliographystyle{iopart-num}
\bibliography{bib}
%\bibitem{bib}

\end{document}